\documentclass[aip,amsmath,amssymb,reprint]{revtex4-1}
\usepackage{epsfig}
\usepackage{amsmath}
\graphicspath{{figures/}}
\usepackage{epsfig}
\usepackage{amsmath}
\usepackage{color}
\usepackage{graphicx,epsfig}
\usepackage{color}
\usepackage[latin9]{inputenc}
\usepackage{float}
\usepackage{natbib}
\begin{document}

%\title{On the electrical conductivities of several NaCl force fields}
\title{On the computation of electrical conductivities of aqueous electrolyte solutions: Two surfaces one property  }

\author{Samuel Blazquez}
\affiliation{Dpto. Qu\'{\i}mica F\'{\i}sica I, Fac. Ciencias Qu\'{\i}micas,
Universidad Complutense de Madrid, 28040 Madrid, Spain.}
\author{Jose Abascal}
\affiliation{Dpto. Qu\'{\i}mica F\'{\i}sica I, Fac. Ciencias Qu\'{\i}micas,
Universidad Complutense de Madrid, 28040 Madrid, Spain.}
\author{Jelle Lagerweij}
\affiliation{Engineering Thermodynamics, Process and Energy Department, Faculty of Mechanical, Maritime and Materials Engineering, Delft University of Technology, Leeghwaterstraat 39, 2628CB, Delft, The Netherlands.}
\author{Parsa Habibi}
\affiliation{Engineering Thermodynamics, Process and Energy Department, Faculty of Mechanical, Maritime and Materials Engineering, Delft University of Technology, Leeghwaterstraat 39, 2628CB, Delft, The Netherlands.}
\affiliation{Department of Materials Science and Engineering, Faculty of Mechanical, Maritime and
Materials Engineering, Delft University of Technology, Mekelweg 2, 2628CD, Delft, The Netherlands}
\author{Poulumi Dey}
\affiliation{Department of Materials Science and Engineering, Faculty of Mechanical, Maritime and
Materials Engineering, Delft University of Technology, Mekelweg 2, 2628CD, Delft, The Netherlands}
\author{Thijs Vlugt}
\affiliation{Engineering Thermodynamics, Process and Energy Department, Faculty of Mechanical, Maritime and Materials Engineering, Delft University of Technology, Leeghwaterstraat 39, 2628CB, Delft, The Netherlands.}
\author{Othonas Moultos}
\affiliation{Engineering Thermodynamics, Process and Energy Department, Faculty of Mechanical, Maritime and Materials Engineering, Delft University of Technology, Leeghwaterstraat 39, 2628CB, Delft, The Netherlands.}
\author{Carlos Vega$^{*}$}
\affiliation{Dpto. Qu\'{\i}mica F\'{\i}sica I, Fac. Ciencias Qu\'{\i}micas,
Universidad Complutense de Madrid, 28040 Madrid, Spain.}

%\date{\today}
%

\begin{abstract}

In this work, we have computed electrical conductivities at ambient conditions of  aqueous NaCl and KCl solutions
	by using the  Einstein-Helfand equation. Common force fields (charge q =$\pm$1 $e$) do not reproduce
	the experimental values of electrical conductivities, viscosities and diffusion coefficients. 
	Recently, we proposed the idea of using different
	charges to describe the Potential Energy Surface (PES) and the Dipole Moment Surface (DMS). 
	In this work, we implement this concept. The equilibrium trajectories required to evaluate 
	electrical conductivities (within linear response theory) 
	were obtained by using scaled charges (with the value $q$ =$\pm$0.75 $e$) to describe the PES. 
The potential parameters were those of the Madrid-Transport force field, which describe accurately viscosities and diffusion coefficients of these ionic solutions. 
	However, integer charges were used to compute the conductivities (thus describing the DMS).  The basic idea is that although the scaled charge
	describes the ion-water interaction better, the integer charge reflects the value of the charge that is transported due to the electric field. 
	The agreement obtained with experiments is excellent, as for the first time electrical conductivities (and the other transport properties) 
	of NaCl and KCl electrolyte solutions
	are described with high accuracy for the whole concentration range up to their
	solubility limit. Finally, we propose an easy way to obtain a rough estimate of the actual electrical conductivity 
	of the potential model under consideration using the approximate Nernst-Einstein equation, which neglects correlations
	between different ions.

\end{abstract}

\maketitle
$^*$Corresponding author: cvega@quim.ucm.es

\section{Introduction}

Electrolyte solutions are ubiquitous in nature, where ions can play a key role in many living organisms\cite{robinson2002electrolyte}. Electrolytes are also important in
many other fields such as battery technology\cite{attias2019anode,logan2020electrolyte,li2020new} and desalination processes\cite{srimuk2020charge}.  
Electrolyte solutions have always been the subject of scientific interest\cite{robinson2002electrolyte,enderby1981structure,barthel1998physical,kirkwood1934theory,pitzer1991activity}
and computer simulations can be a valuable tool for the study of complex phenomena related to these electrolyte solutions in combination with experimental studies.
In the 1970s, Heizinger, Vogel, Singer and
      Sangster\cite{doi:10.1080/00018737600101392,singer77,Heizinger1,heizinger2,Heizinger3}
published the first simulation studies of ionic systems.
However, the computational cost of the simulations and the lack of suitable force fields for water and electrolytes 
did not allow major advances until the recent years.\\

Computational simulations are a useful tool for studying phenomena elusive to experiments and for predicting properties of interest. Nevertheless, 
a suitable force field for the studied system is needed. In the case of aqueous electrolyte solutions, force fields for water and ions are 
necessary\cite{JCP_1933_01_00515}.
In the case of water, the first force field was proposed by Bernal and Fowler in 1933\cite{JCP_1933_01_00515}.
Fifty years later, Jorgensen and coworkers started to develop new potential models, such as the
TIP3P\cite{jorgensen83}, the TIP4P\cite{jorgensen83} and the TIP5P\cite{mahoney01} force fields. At the same time, 
the popular SPC/E force field developed by  Berendsen and coworkers\cite{spce}.
Later, in the 2000s, the knowledge gained from the aforementioned models allowed the development
of two of the water models that best reproduce a wide range of properties, which are TIP4P-Ew\cite{tip4p-ew} and  TIP4P/2005\cite{abascal05b}.
In fact, the TIP4P/2005 potential is able to reproduce a variety of properties such as  densities, viscosities, and the temperature of the maximum in density (TMD)\cite{vega11,tazi2012diffusion,gon:jcp10}.
There is no classical model able to reproduce all properties of pure water\cite{vega11,blazquez2022melting}.
One option to improve results of the previous mentioned force fields is to use polarizable models,
such as the HBP\cite{jiang2016hydrogen}, the MB-Pol\cite{reddy2016accuracy} or the BK3\cite{kissB3k}. However, these force fields are
between three and ten times more computationally expensive than the
non-polarizable force fields and are also not able to reproduce certain properties simultaneously, such as TMD and melting temperature\cite{blazquez2022melting}.
Besides water models, ion-ion and ion-water interaction have to be described to study electrolyte solutions.
It is in the recent years that a large variety of force fields for salts have been
proposed\cite{smith18,opls,str:jcp88,aqvist,dang:jcp92,beg:jcp94,SmithDang,roux:bj96,peng:jcpca97,wee:jcp03,Jorgensen,lam:jcpb06, ale:pre07,len:jcp07,joung08,gallo_anomalies,cal:jpca10,yu:jctc10,reif:jcp11,gee:jctc11,ben:jcp17,deu:jcp12, mao:jcp12,mam:jcp13,mou:jctc13,kiss:jcp14,kol:jcp16,elf:epjst16,pethes17,JCTC_16_2460_2020,JCP_2019_151_134504,madrid_2019_extended,doi:10.1021/acs.jpcb.2c06381,madrid_transport}. 
Among the most popular force fields for ions, we can find the one proposed by
Joung and Cheatham\cite{joung08} (JC),
which includes all the alkali halides. These can be used in combination with three 
different water models-namely; TIP3P, TIP4P-Ew and SPC/E)-the 
JC-SPC/E being the one that provides overall better results.
The other popular force field that is widely used in the literature is the 
one developed by Smith and Dang\cite{SmithDang} (SD) in combination with SPC/E water\cite{spce}.
Although these force fields have been quite successful in describing many properties of electrolytes 
(e.g. densities, structure), they fail in describing properties such as solubilities, viscosities and activity 
coefficients\cite{doi:10.1063/5.0012102}.\\ 

In an attempt to overcome the limitations of current force fields for electrolytes, the idea of using
scaled charges for the ions was suggested.
The concept of scaled charges (i.e., assign a  charge smaller than one for monovalent ions) arises 
from the work of  Leontyev
and Stuchebrukhov\cite{leontyev09,leontyev10a,
leontyev10b,leontyev11,leontyev12,leontyev14}
who proposed a charge of $\pm$0.75 (in electron units) for ions in solution (this was also denoted as Electronic Continuum Correction ECC).
The use of scaled charges has also been proposed by Kann and Skinner \cite{kan:jcp14}. 
However, in this case the value of the scaled charge is selected
in such a way that the potential of mean force between ions at infinite dilution and large distances
should be the same in experiments and in the force field (so that the model recovers
the experimental Debye-Huckel law at infinite dilution). As the potential of mean force depends
on the dielectric constant of the water model, so does the value of the scaled charge leading to a value
of $\pm$0.85 in the particular case of water when described by the TIP4P/2005 model. 
The use of scaled charges for ions in solution has undergone a significant expansion in the last years.
Different groups have proposed new force fields with scaled charges. including those of Jungwirth and 
 coworkers\cite{plu:jpca13,koh:jcpb14,koh:jpcb15,dub:jcpb17,mar:jcp18},
 Barbosa and coworkers\cite{fue:jpc16,fue:pa18}, Li and Wang\cite{li:jcp15} and Bruce and van der Vegt\cite{doi:10.1063/1.5017101}.
 Other authors have proposed the use of different scaled charges for the cation and anion (charging the surrounding water molecules
 to maintain the system electroneutral)\cite{Berkowitz,doi:10.1063/1.4736851,doi:10.1063/1.3589419,doi:10.1063/1.4874256,SONIAT201631}.
Breton and Joly also studied the effect of introducing scaled charges for studying interfacial properties\cite{joly_st}.
In this context, we developed a model for NaCl based on scaled charges\cite{ben:jcp17}. Later, we considered a larger  number
of salts and we proposed the Madrid-2019 force field\cite{JCP_2019_151_134504,madrid_2019_extended}
which includes all the possible alkali halides, some divalent salts (Mg$^{2+}$ and Ca$^{2+}$)
and sulphates (SO$_4^{2-}$). We have shown in previous works that this force field is able to reproduce
different properties of interest, such as the salting out effect of methane\cite{FPE_2020_513_112548}, the TMD of different salt solutions\cite{sedano2022maximum},
the freezing depression of ice in presence of different electrolytes\cite{doi:10.1063/5.0085051} or different properties
of seawater\cite{doi:10.1021/acs.jctc.1c00072}.
Although scaled charges improve the results in the majority of properties with respect to unit charge models,
there is no unique value of the scaled charge that describes all properties correctly.
As we have recently shown, the scaled charge can be taken as a fitting parameter depending on the property that one wants to reproduce\cite{madrid_transport}.
Transport properties are among the most interesting properties that can be studied by simulation and 
that traditional force fields of electrolytes have never been able to reproduce correctly.\cite{kim:jpcb12}. In our recent work, we proposed
the Madrid-Transport force field\cite{madrid_transport,doi:10.1021/acs.jpcb.2c06381}, which uses an scaled charge of $q$=$\pm$0.75 and that is able to reproduce transport
properties, such as the viscosities and diffusion coefficients of water and ions in the whole concentration range. This force field
has also been able to reproduce transport properties in the presence of hydrogen\cite{van2023interfacial}.
It should be mentioned that the introduction of scaled charges, improves a number of properties
of electrolytes but deteriorates the value of the free energy of solvation (although it can be corrected via theoretical
corrections\cite{FPE_2020_513_112548}).\\

In this work, we want to analyze in detail the quality of the predictions for electrical conductivities of force fields
using either integer or scaled charges. To the best of our knowledge, such a detailed comparison has never been presented before.
As we will show in the next section, electrical conductivities can be 
calculated with the Einstein-Helfand (EH) 
equation\cite{schroder2008computation,picalek2007molecular,nieszporek2016calculations,helfand1960transport,malaspina2023transport,celebi2019structural,dawass2022solubilities}. 
In a preliminary but pioneering work, Lyubartsev and Laaksonen\cite{lyubartsev1996concentration} evaluated the electrical conductivities of 
NaCl aqueous solutions at different concentrations with the flexible SPC model for water and ions described as charged LJ particles by using the 
Green-Kubo (GK) equation\cite{allen_book} (which is strictly equivalent to the Einstein-Helfand relation).
Due to the computational cost of evaluating the conductivities in this way\cite{kubisiak2020estimates}, many authors 
tend to calculate the electrical conductivities by using the Nernst-Einstein equation 
(i.e., neglecting the ion-ion correlations)\cite{loche2021transferable,hu2010assessment,yllo2019experimental,zhang2015direct,krienke2007hydration,prasad2017concentration}.
Although this is cheaper from a computational point of view, the results are not exact (as this is an approximation)
and overestimate the real conductivities of the model. 
Electrical conductivities have been accurate and extensively calculated (through GK or EH) by different authors for 
ionic liquids\cite{picalek2007molecular,kowsari2009molecular,kowsari2011molecular,mondal2014molecular,rey2006transport,schroder2008computation}.
In the case of aqueous electrolyte solutions, there are a few studies in which the conductivities were properly evaluated\cite{nieszporek2016calculations,sala2010effects,shao2020role}. 
Mart\'{i}, Guardia and coworkers\cite{sala2010effects} calculated electrical conductivities of NaCl solutions at different concentrations using the 
Smith and Dang\cite{SmithDang} ion force field with SPC/E water\cite{spce} (SD-SPC/E). Shao et \textit{al.}\cite{shao2020role} also calculated 
the conductivities of NaCl solutions but using in this case the Joung and Cheatham\cite{joung08} force field in combination with SPC/E water (JC-SPC/E). In fact,
these authors showed interesting results about the existence of finite size effects when calculating conductivities with the Nernst-Einstein relation.
There are also semi-empirical fitted models which try to reproduce the electrical conductivities of NaCl solutions in solvent mixtures, such as 
water-propylene carbonate\cite{zhang2020experimental} or water-monoethylene glycol\cite{moura2021density}. Other authors have also rigorously computed
electrical conductivities for molten salts\cite{grasselli2019topological,pegolo2020oxidation}.
Nevertheless, there is no comprehensive Molecular Dynamics (MD) study of the performance of different force fields of ions and water
for reproducing the electrical conductivities of NaCl solutions in water.

The main purpose of this work is to provide a benchmark to calculate electrical conductivities of aqueous electrolyte solutions,
to show that there is a force field able to reproduce the experimental conductivities of NaCl and KCl solutions and finally, 
from a deeper perspective, we want to demonstrate that to reproduce conductivities the two surfaces present in water have to be 
simultaneously described.
We will properly evaluate the electrical conductivities of different well-known
ion force fields in combination with different water force fields. 
Besides, in this work, we shall introduce a new "conceptual" strategy to determine electrical conductivities.
As we have mentioned in the past,
water (and aqueous solutions) has two surfaces:
the potential energy surface (PES) that describes the energy of each configuration of the system and the dipole moment surface (DMS) 
that describe the dipole moment of each configuration\cite{vegamp15,predota-trampacargasgromcas,jorge2019dielectric}.
The key idea is that these two surfaces can be described by different fitting parameters (i.e., with different charges in our case).
In this work, we use a force field with scaled charges to perform the Molecular Dynamics simulations
of the aqueous electrolyte solutions.
Only the PES is needed to perform these simulations as we are using the
linear response theory, that evaluates transport properties (electrical
conductivities in this work) by analyzing the fluctuations of the 
system when at equilibrium (in absence of an electric field in the case
of electrical conductivities). We will use the Madrid-Transport force
field ($q$=$\pm$0.75) for the PES as it is able to describe other properties
of electrolyte solutions accurately (i.e., viscosities and diffusion coefficients).
The obtained trajectories  are analyzed employing unit charges to calculate the electrical conductivities
(as integer charges instead of partial charges provide better representation of
the DMS).
In this way, we will show how to evaluate conductivities of aqueous electrolyte solutions with a new methodology which yields to 
reproduce the experimental conductivities of NaCl and KCl solutions by using a scaled charge force field.

%we will study the JC-SPC/E and SD/SPC/E force fields (both using SPC/E
%water and with unitary charges for the ions). We will also study the effect of the charge of the ions by using the Madrid-2019 force field with scaled charges and the 
%TIP4P/2005 water model (scaled charges for the trajectories and unit charges for the conductivities).
%Finally we will examine the recent proposed Madrid-Transport force field which provides excellent results for other transport
%properties such as viscosities and diffusion coefficients.

\section{Methodology}

%Conductivities can be calculated using the Green-Kubo (GK) equation\cite{allen_book} as described in Eq. \ref{conduc-GK}
%\begin{equation}
%\label{conduc-GK}
%        \sigma = \frac{V}{k_{B}T}  \int_{0}^{\infty} \Big \langle \Vec{j}_{i}(t)\cdot\Vec{j}_{i}(t_{0}) \Big \rangle dt ,
%\end{equation}
%where V is the system volume, k$_{B}$ the Boltzmann constant and $\Vec{j}_{i}$ is the current density vector which is defined
%in Eq. \ref{current}
%
%\begin{equation}
%\label{current}
%	\Vec{j}_{i}(t)=q_{i} \cdot \Vec{v}_{i}(t),
%\end{equation}
%
%where q$_i$ and $\Vec{j}_{i}$ are the charge and velocity (at a certain time)  of the i$-th$
%particle respectively.
%Eq. \ref{conduc-GK} can be turned into the equivalent 

Conductivities can be calculated using the Einstein-Helfand relation\cite{schroder2008computation,picalek2007molecular,nieszporek2016calculations,helfand1960transport,malaspina2023transport}:

\begin{equation}
\label{conduc-einstein}
	\sigma = \lim_{t \to \infty} \frac{\mathrm{d}}{\mathrm{d}t} \frac{1}{6Vk_{\mathrm{B}}T}  \Big \langle \left\lbrace \sum_{i}^{N} q_{i}[\Vec{r}_{i}(t)-\Vec{r}_{i}(0)] \right\rbrace ^{2} \Big \rangle ,
\end{equation}
where $V$ is the system volume, $k_\mathrm{{B}}$ the Boltzmann constant, $T$ the temperature, $\Vec{r}_{i}(t)$ and $\Vec{r}_{i}(0)$  the position of the $i^{th}$
particle at time $t$, and the $\langle
[\Vec{r}_{i}(t)-\Vec{r}_{i}(0)]^{2}\rangle$ term is the mean square displacement (MSD).
Taking into account that the dipole moment of the system can be defined as:

\begin{equation}
\label{dipole}
	\Vec{M}(t) = \sum_{i}^{N} q_{i}\cdot\Vec{r}_{i}(t)
\end{equation}
we can obtain the following equation for the conductivity in function of the mean square dipole displacement 
of the system:

\begin{equation}
\label{conduc-einstein2}
	\sigma = \lim_{t \to \infty}  \frac{\mathrm{d}}{\mathrm{d}t} \frac{1}{6Vk_{\mathrm{B}}T}  \Big \langle [M(t)-M(0)]^{2} \Big \rangle ,
\end{equation}
This way, the conductivity can be easily obtained from the slope of the mean square dipole displacement
versus time (in the Supporting Information we show this
plot for different initial seeds of the MD simulations).\\
%Note here that our approach is different than previous performed studies about the same problem. 
%As we have previously pointed out,
%water has two surfaces:
%The potential energy surface (PES) and the dipole moment surface (DMS)\cite{vegamp15,predota-trampacargasgromcas}.
%In this work we have performed the simulations with the different force fields and their respective charges. Nevertheless we have analyzed the
%trajectories employing unit charges to calculate the electrical conductivities
%(i.e., the formal charges employed in Eq. \ref{dipole} have been always $\pm1$).
%Thus, in the case of the scaled charges models, the simulation have been performed with the scaled charges (i.e., $\pm$0.75 and 
%$\pm$0.85 for the Madrid-Transport and Madrid-2019 respectively) but afterwards, the formal charge for the analysis (Eqs. \ref{conduc-einstein}-\ref{conduc-einstein2})
%is $\pm$1.\\

Eq. \ref{conduc-einstein} can be rewritten as:

\begin{widetext}
\begin{equation}
\label{conduc-delftdemost}
	\sigma = \lim_{t \to \infty}  \frac{\mathrm{d}}{\mathrm{d}t} \frac{e^2}{6Vk_{\mathrm{B}}T}  \Big \langle  \sum_{i}^{N} \sum_{j}^{N} z_{i}z_{j}[\Vec{r}_{i}(t)-\Vec{r}_{i}(0)][\Vec{r}_{j}(t)-\Vec{r}_{j}(0)]  \Big \rangle ,
\end{equation}
\end{widetext}
Please note that  $q_i$ is equivalent to $e$$\cdot$z$_i$
We define particles between 1 and $\frac{N}{2}$ as cations, and particles between $\frac{N}{2} +1$ and $N$ as anions. Thus, we can define for a 1:1 electrolyte the 
Onsager coefficients\cite{robinson2002electrolyte} ($\Lambda_{ij}$) for the different interactions between cations (+) and and anions (-):

\begin{equation}
\label{onsagerplusplus}
	\Lambda_{++} = \lim_{t \to \infty}  \frac{\mathrm{d}}{\mathrm{d}t} \frac{1}{6N}  \Big \langle  \sum_{i=1}^{\frac{N}{2}} \sum_{j=1}^{\frac{N}{2}} [\Vec{r}_{i}(t)-\Vec{r}_{i}(0)][\Vec{r}_{j}(t)-\Vec{r}_{j}(0)]  \Big \rangle ,
\end{equation}

\begin{equation}
\label{onsagerplusminus}
        \Lambda_{+-} = \lim_{t \to \infty}  \frac{\mathrm{d}}{\mathrm{d}t} \frac{1}{6N}  \Big \langle  \sum_{i=1}^{\frac{N}{2}} \sum_{j=\frac{N}{2}+1}^{N} [\Vec{r}_{i}(t)-\Vec{r}_{i}(0)][\Vec{r}_{j}(t)-\Vec{r}_{j}(0)]  \Big \rangle ,
\end{equation}

\begin{equation}
\label{onsagerminusplus}
        \Lambda_{-+} = \lim_{t \to \infty}  \frac{\mathrm{d}}{\mathrm{d}t} \frac{1}{6N}  \Big \langle  \sum_{i=\frac{N}{2}+1}^{N} \sum_{j=1}^{\frac{N}{2}} [\Vec{r}_{i}(t)-\Vec{r}_{i}(0)][\Vec{r}_{j}(t)-\Vec{r}_{j}(0)]  \Big \rangle ,
\end{equation}

\begin{equation}
\label{onsagerminusminus}
        \Lambda_{--} = \lim_{t \to \infty}  \frac{\mathrm{d}}{\mathrm{d}t} \frac{1}{6N} \Big \langle  \sum_{i=\frac{N}{2}+1}^{N} \sum_{j=\frac{N}{2}+1}^{N} [\Vec{r}_{i}(t)-\Vec{r}_{i}(0)][\Vec{r}_{j}(t)-\Vec{r}_{j}(0)]  \Big \rangle ,
\end{equation}
where $N$ is the total number of ions.
Combining Eq. \ref{conduc-delftdemost} with Eqs. \ref{onsagerplusplus}-\ref{onsagerminusminus}, 
the electrical conductivity can be calculated from:

\begin{equation}
	\sigma = \frac{e^{2}N}{Vk_{\rm{B}}T} \sum_{i,j}z_iz_j\Lambda_{ij}
    \label{sigma-delft}
\end{equation}

Eq. \ref{sigma-delft} is strictly equivalent to Eqs. \ref{conduc-einstein} and \ref{conduc-einstein2}.
Note that in all these equations, we only include the charge and positions of the ions. We do not consider
water (solvent) molecules, because it is a neutral molecule which does not contribute to the electrical conductivity.\\

Since the evaluation of electrical conductivities by the previous methodology can be computationally demanding, some authors use the 
approximate 
Nernst-Einstein (NE)
equation\cite{einstein1905molekularkinetischen}, which relates the electrical conductivity to the self-diffusion
coefficients of the ions. 
\begin{equation}
\label{nernst-einstein}
        \sigma = \frac{q^{2}\rho}{k_{\mathrm{B}}T}(D_{+}+D_{-}),
\end{equation}
where $q$ is the charge of the ions, $\rho$ is the  number density of the salt and $D_{+}$ and $D_{-}$ are the self-diffusion coefficients of the
cation, and anion respectively.
The NE approximation assumes 
that the only terms which contribute to the conductivities are those of the particle with itself.
Note that whereas there are $N^2$ terms in the rigorous expression of the conductivity (Eq. \ref{conduc-delftdemost}), 
there are only $2N$ terms when using the NE relation (Eq. \ref{conduc-delftdemost}).
The NE relation, thus, can be obtained from Eq. \ref{conduc-einstein} by
assuming that the displacement of ions is independent (i.e.,  $\langle r_{i} \cdot r_{j}\rangle$=0 for $i$$\neq$$j$).
Making this assumption, electrical conductivities can be easily obtained for equimolar salts with Eq. \ref{nernst-einstein}.
For the evaluation of conductivities with the NE equation, we have calculated the self-diffusion coefficients of both Na$^+$ and Cl$^-$
by using the Einstein relation (Eq. \ref{eq_diffus}):

\begin{equation}
\label{eq_diffus}
	D_{i}^{\mathrm{MD}} = \lim_{t \to \infty} \frac{1}{6 t}  \Big \langle [\Vec{r}_{i}(t)-\Vec{r}_{i}(0)]^{2} \Big \rangle ,
\end{equation}
where $\Vec{r}_{i}(t)$ and $\Vec{r}_{i}(0)$ are the position of the $i^{th}$
particle at time $t$ and 0, and the $\langle
[\Vec{r}_{i}(t)-\Vec{r}_{i}(0)]^{2}\rangle$ term is the MSD.
All diffusivities in this work are corrected using  
the hydrodynamic corrections of Yeh and Hummer\cite{Yehhummer,celebi2021finite} which are
described as
\begin{equation}
\label{corectionyeh}
	D_{i} = D_{i}^{\mathrm{MD}} + \frac {k_{\mathrm{B}}T\xi}{6\pi \eta L}
\end{equation}
where $D_{i}$ is the diffusion coefficient with the applied corrections of Yeh and Hummer,
$D_{i}^{\mathrm{MD}}$ is the diffusion coefficient initially obtained by simulations,
$\xi$ is a dimensionless constant equal to 2.837,
$\eta$ is the computed  viscosity at the studied concentration (which is shown to exhibit no finite size effects\cite{moultos2016system,jamali2018shear,Yehhummer}), and $L$
is the length of the simulation box.

\section{Simulation Details}

Electrical conductivity is not a property for which there are many previous results and it is interesting to see 
whether two groups using different software packages (GROMACS vs LAMMPS), slightly different methodologies 
(the individual calculation of the Onsager coefficients $versus$ the global expression), and  different system sizes yield the same result for a specific model. 
Since we also aim to use the results of this work as a benchmark for people computing electrical conductivities in the future, 
and we want to ensure  that the good agreement with experiment that is observed here is true regardless 
of the MD software or post-processing program used for the computation. 
As one group is located in Madrid (UCM) and the other in Delft (TU Delft), they will be denoted as Madrid and Delft groups.
Concentrations of the salts will be given in molality units, so that a solution with a concentration 1 m corresponds to 
1 mol of salt per kilogram of water. 
All the results of this work correspond to room temperature and atmospheric pressure (i.e., 298.15 K and 1 bar).
Note also that error bars of all the results are calculated by both groups as the
standard deviation the property obtained in each run using different initial seeds divided by the square root of the number of runs.

\subsection{Madrid Group}
All MD simulations were performed with GROMACS\cite{spoel05,hess08} (version 4.6.7).
The leap-frog integrator algorithm\cite{bee:jcp76} with a time step of 2 fs was used.
We have also employed 
periodic boundary conditions in all directions.
Temperature and pressure were kept constant using the Nos\'e-Hoover thermostat\cite{nose84,hoover85} 
and Parrinello-Rahman barostat\cite{parrinello81}, both with a coupling constant of 2 ps.
For electrostatics and van der Waals interactions, the cut-off radii were fixed at 1.0 nm and long-range
corrections in the energy and pressure were applied to the Lennard-Jones part of the potential.
The smooth PME method\cite{essmann95} was used to account for
long-range electrostatic forces.
Water geometry was maintained using
the LINCS algorithm\cite{hess97,hess08b}.
To compute conductivities, we have simulated systems of 4440 water molecules and the corresponding  number of ions for the
desired concentration (e.g., 80 NaCl molecules for a concentration of 1 $m$ as shown in Tables \ref{tab:NaCl_table} and  \ref{tab:KCl_table}). We have performed
an initial  $NpT$ simulation of 20 ns to accurately
calculate the volume of the system. After that, using the average volume obtained in the $NpT$ simulation
we have carried out five independent runs in $NVT$ ensemble. Runs of 200 ns were performed for the lowest
concentration (i.e., 1 m) and of 120 ns for the higher concentrations (2, 4, and 6 m). Thus, typically around 600 ns (5x120) or 1 $\mu$s (5x200)
are needed to compute electrical conductivities of each model and thermodynamic state.
The electrical conductivities were obtained from fitting the mean square dipole displacement (Eq. \ref{dipole})
versus time between 50 and 1000 ps as shown in Eq. \ref{conduc-einstein2} (in the SI we also provide the results from fitting the data between 50 and 2000 ps).

\subsection{Delft Group}

MD simulations are carried out
on the DelftBlue super-computer at TU Delft\cite{DHPC2022}
with the Large-scale Atomic/Molecular Massively Parallel Simulator (LAMMPS: version August 2018)~\cite{Plimpton1995}.
Periodic boundary conditions are imposed in all directions, and the velocity-Verlet algorithm is used with a time step of 2 fs.
The Nos\'e-Hoover thermostat and barostat~\cite{nose84,hoover85,parrinello81} are set with coupling constants of 0.1 and 1 ps, respectively.
The SHAKE algorithm is used to fix the bond lengths and angles of water~\cite{Plimpton1995,Ryckaert1977}.
A cut-off of 1.0 nm is used for both the Lennard-Jones and electrostatic potentials. Long-range electrostatic interactions are modelled
using the particle-particle particle-mesh (PPPM) method~\cite{Frenkel1996, Hockney2021} with a relative error~\cite{LAMMPSMANUAL} of $10^{-5}$. 
Analytic tail corrections\cite{Frenkel1996} are applied to the Lennard-Jones interactions for both energies and pressures. 
Initial configurations are created using PACKMOL (v20.3.1)~\cite{Martinez2009}.
To compute electrical conductivities, self-diffusivities, and shear viscosities, the OCTP plugin~\cite{Jamali2019} is used. In this plugin,
the Einstein relations are used in combination with the order-$n$ algorithm\cite{Frenkel1996, Dubbeldam2011} to compute transport properties.
All details on the OCTP plugin can be found in Ref.\cite{Jamali2019}
The approach to evaluate conductivities is based on the computation of the Onsager coefficients ($\Lambda_{ij}$)
for the cation-cation, anion-anion and cation-anion interactions independently as shown in Eqs. \ref{conduc-delftdemost}-\ref{sigma-delft}. 
These equations are used to compute the exact electrical conductivities accounting for ion-ion correlations.
In the SI we have collected the results for the individual contributions to the electrical conductivities (i.e., $\sigma$$_{++}$,$\sigma$$_{+-}$, $\sigma$$_{-+}$, and $\sigma$$_{--}$)
computed from each individual Onsager coefficient.
To evaluate the electrical conductivities, the system sizes were of 555 or 1000 water molecules (the corresponding number of ion molecules is dictated from the molality
of each system) as we have listed in
Tables \ref{tab:NaCl_table} and  \ref{tab:KCl_table}.
To accurately compute the average volume of the simulation box, simulations of 20 ns in the $NpT$ ensemble were initially carried out 
(10 ns equilibration runs followed by 10 ns production runs). The self-diffusivities, 
viscosities, and Onsager coefficients are calculated from production runs of 200 ns in the $NVT$ ensemble. 
Three different simulations were carried out with different initial velocities for all molalities to obtain statistics.
Thus, a total simulation time of  ca. 600 ns
is required to compute electrical conductivities of each model and thermodynamic state.

%\begin{equation}
%    \sigma = \lim_{t\to \infty}\frac{e^2}{6tVk_{\rm{B}}T}\sum_{\it{i,j}}\sum_{k=1}^{N_{i\color{white}{j}}}\sum_{l=1}^{N_j}\it{z_i z_j} \langle [\bf{r}_{\it{k,i}}\it{(t)}-\bf{r}_{\it{k,i}}\rm{(0)}][\bf{r}_{\it{l,j}}\it{(t)}-\bf{r}_{\it{l,j}}\rm{(0)}] \rangle
%    \label{Exact Econd}
%\end{equation}
%
%where $\sigma$ is the electrical conductivity, $e$ is the elementary charge, $t$ is time, $V$ is the volume of the simulation box, and $N_i$ is the number of ions of type $i$. $z_i$ and $\bf{r}_{\it{k,i}}\it{(t)}$ represent the charge of ion $i$ and the position vector of the $k-$th ion of type $i$ at time $t$, respectively.
%
%\begin{equation}
%    \Lambda_{ij} = \frac{1}{N}\lim_{t\to\infty} \frac{1}{6t} \Bigg \langle \sum_{k=1}^{N_{i\color{white}{j}}}\sum_{l=1}^{N_j}[\bf{r}_{\it{k,i}}\it{(t)} \bf-{r}_{\it{k,i}}\rm{(0)}] [\bf{r}_{\it{l,j}}\it{(t)} \bf-{r}_{\it{l,j}}\rm{(0)}]  \Bigg \rangle
%    \label{Ons}
%\end{equation}
%
%where $N$ is the total number of molecules. Combining Eqs.~\ref{Exact Econd} and \ref{Ons}, it can be derived that \cite{Fong2020}:
%
%\begin{equation}
%    \sigma = \frac{e^2 N}{Vk_{\rm{B}}T} \sum_{\it{i,j}} z_i z_j \Lambda_{ij}
%    \label{sigma}
%\end{equation}
%

\section{Results}

\subsection{Electrical conductivities of popular force fields}

\begin{center}
\begin{figure}[!hbt] \centering
    \centering
    \includegraphics*[clip,scale=0.35,angle=0.0]{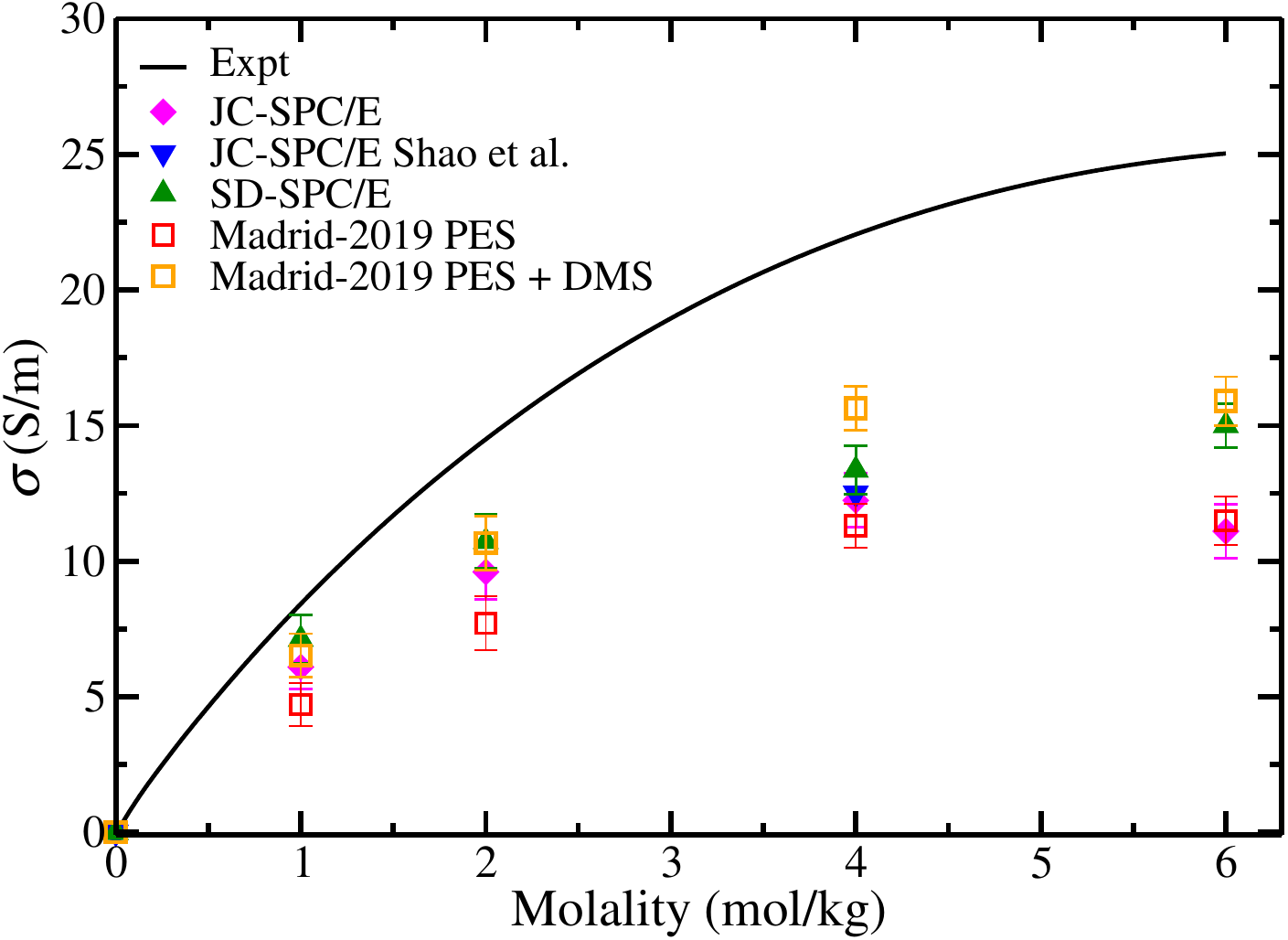}
	\caption{Electrical conductivities (computed by the Madrid group) as a function of NaCl concentration
	obtained with the different models studied in this work using Eq. \ref{conduc-einstein2} at temperature $T$ = 298.15 K, and pressure $p$ = 1 bar. 
Note that Madrid-2019 PES (red empty squares) uses scaled charges in both MD simulations and for computing the conductivities, but Madrid-2019 PES+DMS uses 
scaled charges for the dynamics and integer charges for computing the conductivities.
	Experimental results have been
	taken from Ref.\cite{chambers1956conductances}}
    \label{conductivities-unit}
\end{figure}
\end{center}

In previous studies, we have shown that popular force fields (that use integer charges for the ions) are not able to reproduce transport properties,
such as viscosities, diffusion coefficients of water in salt solutions, and self-diffusion coefficients of ions\cite{madrid_transport,JCP_2019_151_134504,madrid_2019_extended}. 
Here we investigate if these models also fail in accurately predicting another important transport property, 
namely the electrical conductivity. 
In Fig. \ref{conductivities-unit}, results of the electrical conductivity of NaCl  obtained in this work for two popular force fields 
(i.e., Joung and Cheatham\cite{joung08} and Smith and Dang\cite{SmithDang}) that use integer charges combined with the SPC/E water model 
are compared to experiments. 
Note that contrary to previous studies that only focused on concentrations up to 4 $m$, in this work we have evaluated 
the conductivities in the whole concentration
range, i.e., up to the experimental solubility limit of NaCl (6.1 $m$).
Our results at 4 $m$ for the JC-SPC/E are in excellent agreement with those obtained by Shao et \textit{al.}\cite{shao2020role}
using the Green Kubo formalism.
Results  presented in Fig. \ref{conductivities-unit} were obtained by the Madrid group. 
It is clear that electrical conductivities are underestimated with respect to experiments
by the models using integer values for the charge. 
The SD-SPC/E is slightly more accurate compared to the JC-SPC/E,
but also underestimates the electrical conductivities.  
Thus, it is evident that these two popular force fields for NaCl are not able to reproduce electrical conductivities. These models overestimate
the experimental viscosities of NaCl solutions\cite{madrid_transport}. 
It is not surprising that electrical conductivities are underestimated,
as intuitively one would expect that an overestimate of the viscosity would lead 
to an underestimate of the diffusion coefficient of the ions and, therefore, to an underestimate of the the electrical conductivity.

\subsection{Electrical Conductivities of Scaled Charge models}

Unit charge force fields cannot reproduce electrical conductivities of aqueous solutions, but in previous works we have demonstrated that scaled charge models significantly improve the
description of transport properties as viscosities and self-diffusion coefficients\cite{madrid_transport,doi:10.1021/acs.jpcb.2c06381}. Thus, we have now employed the Madrid-2019
force field to observe whether we can improve the results of unit charge force fields. This force field uses scaled charges 
for the ions (in particular 0.85). In Fig. \ref{conductivities-unit}, we show the results for the conductivities of the Madrid-2019 
model (red empty squares). Surprisingly we do not obtain better results than those of unit charge models. 
This was also observed by Gullbrekken \textit{et al.}\cite{gullbrekken2023charge} in their recent work in which they studied the electrical conductivities
by using an integer charge force field but then when scaling the charges the do not observe differences.
Regarding Madrid-2019 results, one may wonder
how is it possible than a force field which yields better results for viscosities and diffusion 
coefficients of aqueous solutions
does not perform equally well for electrical conductivity.
We shall provide now a possible explanation for this puzzling behavior. As discussed in detail previously\cite{vegamp15}, water has two different surfaces, the Potential Energy Surface (PES) and 
the Dipole Moment Surface (DMS).
In the absence of an external macroscopic  electric field, all properties of a system can be determined in computer simulations from the PES.
The DMS surface is not needed to determine any property of a system when not applying an external electric field. 
However,  certain properties
describe the response of a system to an external electric field.
In particular, the dielectric constant and the electrical conductivity are response functions of this type.
Obviously, these properties are relevant only when an external electric field is applied to the system.
Therefore, to determine these response functions in computer simulations, it is required to describe both the PES and the DMS.
A clarification is in order now. The PES simply gives the energy of a system provided the positions of all the nuclei of the system. 
The PES only depends on the position of the atoms, and does not depend on any macroscopic property as for instance the viscosity, or the dielectric constant. 
Sometimes it is stated that the dielectric constant enters in the description of the PES in the case of electrolytes. This is not correct. 
Even for electrolytes one simply needs to know the position of the atoms to determine the energy of the system, and the value of the dielectric constant 
is not needed. The origin of this confusion arises from the fact that at infinite dilution, and when $r$ tend to $\infty$, the potential of mean 
force $w(r)$ between two ions obtained defined from the radial distribution function $g(r)$ using the expression:
\begin{equation}
\label{gder-wder}
	g(r)=e^{-\beta w(r)},
\end{equation}
can be obtained from the knowledge of the dielectric constant $\epsilon_r$ as follows:
\begin{equation}
\label{wder}
	w(r) = \lim_{\substack{ r \to \infty \\ m \to 0 }} \frac{q_1 q_2}{ 4\pi \epsilon_{0} \epsilon_r  r},
\end{equation}

But the potential of mean force is not the PES , and besides this expression is only valid in the Debye-Huckel limit (i.e., at infinite dilution of electrolyte 
  and infinitely large distances). The summary is that the dielectric constant does not enter in the description of the PES, and in the absence of 
  an external electric field all properties of an aqueous electrolyte solution can be obtained from the PES and the knowledge of the DMS is not needed.

Imagine that a model is able to describe correctly the viscosities and the diffusion coefficients.  This is an indicator that the PES is described correctly.
Imagine now that the electrical conductivity is not well described. How to solve this paradox?
The answer is rather simple: the PES is well described but the DMS is not well described. 
  Often partial charges are used to describe the PES and these charges are also used to describe the DMS.
  We have suggested sometime ago that the charges that are good to describe the PES may not be suitable to describe the DMS\cite{vegamp15}.
  We have provided some indications by analyzing the behavior of the dielectric constant of water.  Jorge and coworkers\cite{jorge2019dielectric,jorge2023optimal}
  have shown that the same is true for the dielectric constant of alcohols, and Bowman and coworkers followed up on this idea\cite{liu2016transferable,liu2015quantum}.

\begin{center}
\begin{figure*}[!hbt] \centering
    \centering
    \includegraphics*[clip,scale=0.31,angle=0.0]{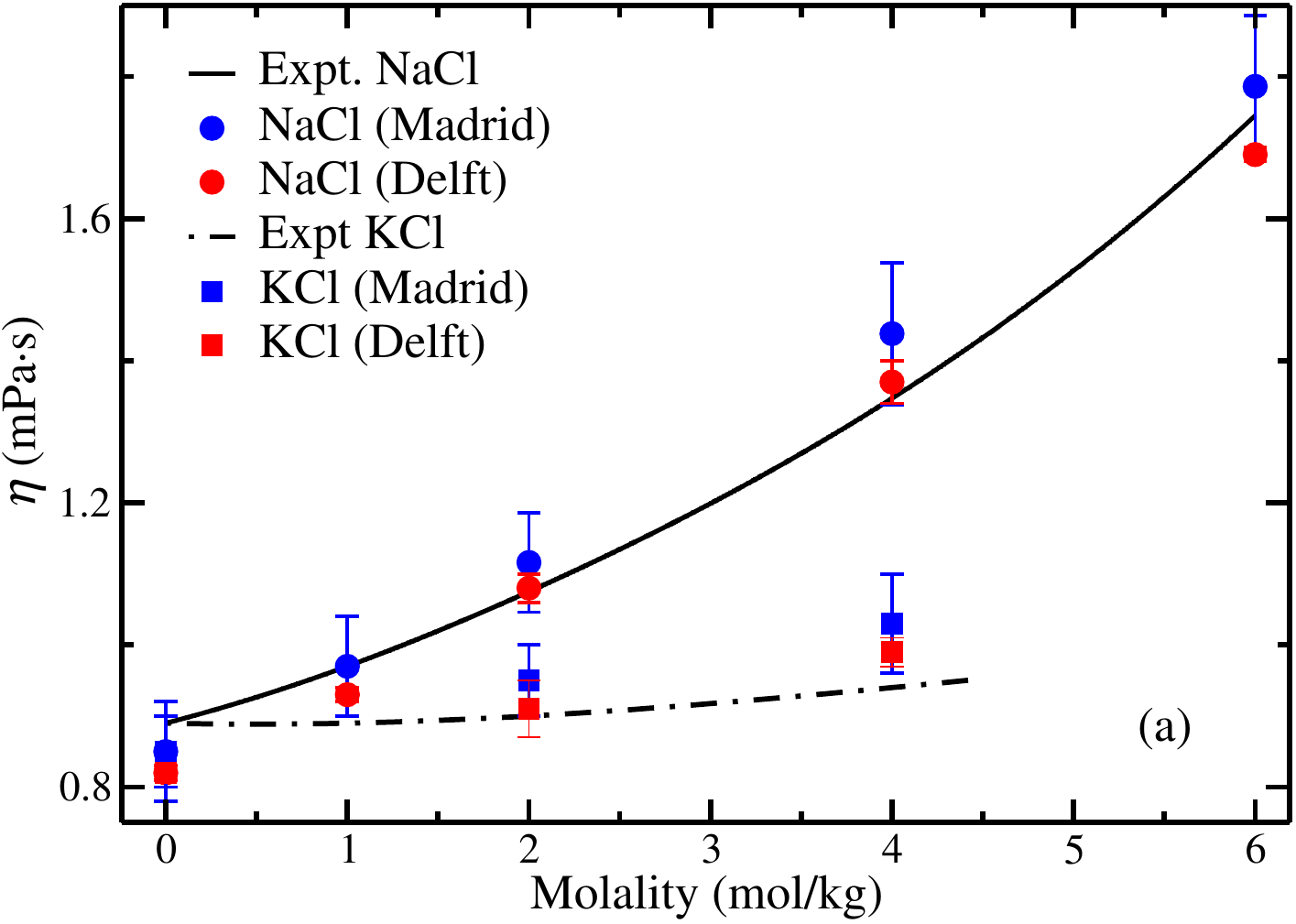}
    \includegraphics*[clip,scale=0.31,angle=0.0]{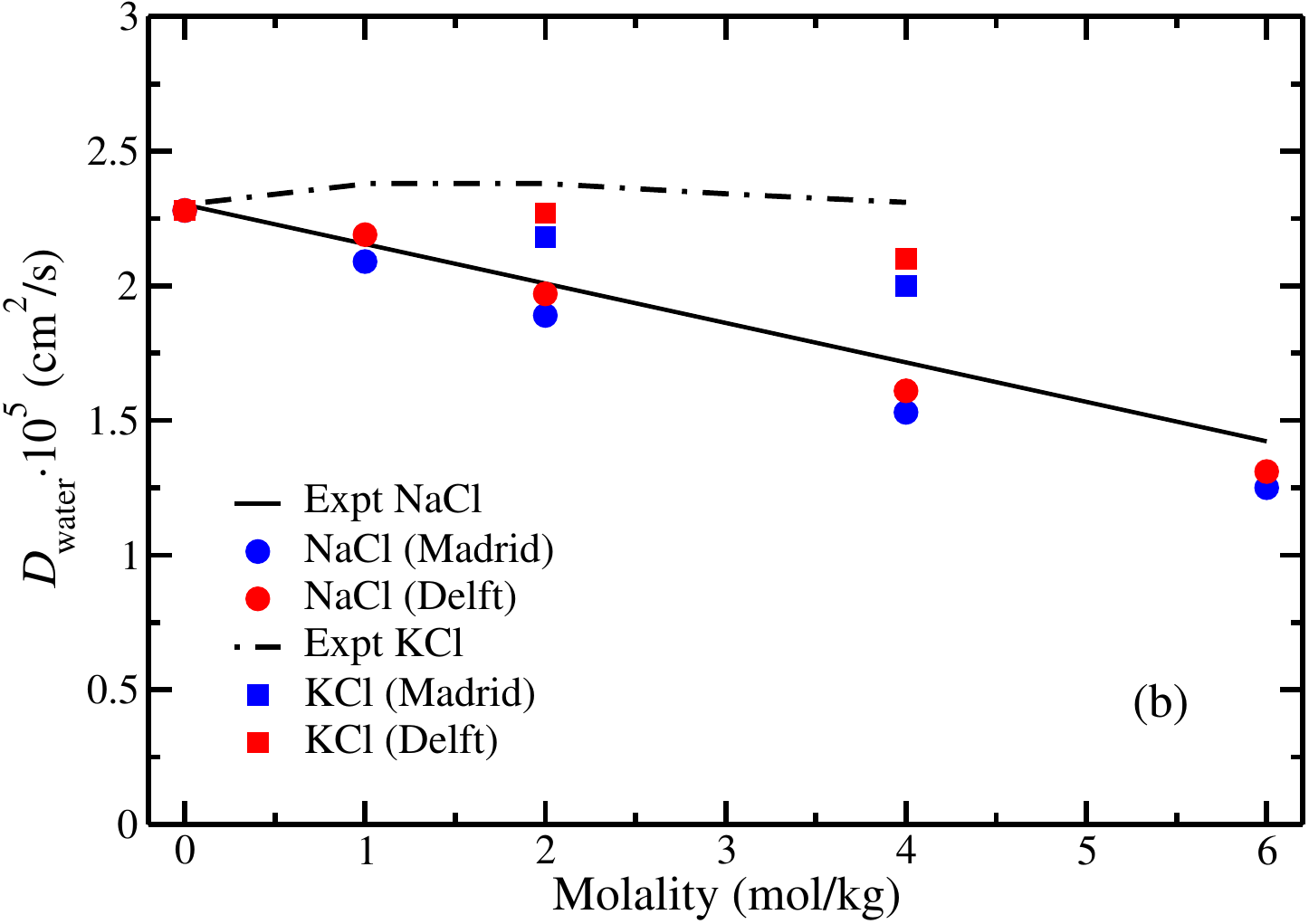}
        \caption{Transport properties of  NaCl and KCl aqueous solutions at different concentrations
	obtained with the Madrid-Transport force field by Madrid (blue) and Delft (red) groups at at temperature $T$ = 298.15 K and pressure $p$ = 1 bar.
	(a) Viscosities and (b) Self-diffusion coefficients of water (corrected for system size effects).
	Experimental results have been
        taken from Refs.\cite{lal:jced07,mul:jpc96}}
    \label{transport-properties}
\end{figure*}
\end{center}

  Here, we shall use different charges to describe the PES and the DMS of electrolytes in water.
  In particular, we shall use scaled charges for the ions in water for the PES while we shall use
  non-scaled charges (i.e., integer charges) to describe the DMS.
The way to implement this idea is rather simple. Since we are using linear response theory 
(so that the electrical conductivity is computed by simulating the system in the absence of the electric field),
we shall perform MD simulations to obtain the trajectories using the scaled charge of the ions  (i.e., $\pm$ 0.85 for the particular case of the 
Madrid-2019 force field).  
In sharp contrast, when the trajectory is analyzed using Eqs. \ref{conduc-einstein}-\ref{sigma-delft}, integer charges are used for the ions.
The results from this approach are presented for the 
Madrid-2019 force field in Fig. \ref{conductivities-unit} (orange empty squares). 
As clearly shown, the computed conductivitiesare closer to the experimental data.
These results showcase that a better description is obtained when simultaneously describing both
surfaces the PES and DMS. However, we do not obtain a perfect agreement with the experimental electrical conductivities of the aqueous solutions.
This is not entirely surprising, as the Madrid-2019 force field (that uses a scaled charge of $\pm$0.85)  improves the description
of transport properties of electrolytes in water, but is not able to yield quantitative  agreement with experiments. 
  To this end, we have recently proposed a force field for NaCl and KCl (denoted as Madrid-Transport) that is able
  to predict transport properties of these electrolyte solutions with excellent agreement to experiment\cite{madrid_transport,doi:10.1021/acs.jpcb.2c06381}.
  Therefore, we shall compute the electrical conductivities of NaCl and KCl solutions using the Madrid-Transport force field (that uses an scaled charge of $\pm$0.75).
  We shall implement the main idea of this work, namely, to use scaled charges to obtain the trajectories and integer charges to describe the DMS (i.e.,
  using integer charges in Eqs. \ref{conduc-einstein}-\ref{sigma-delft}). 
In Fig. \ref{transport-properties}, we present both properties (viscosities and self-diffusion coefficients of water) in the whole concentration
range of each salt up to the experimental solubility limit. Results are independently calculated by two different research groups: Madrid (blue) and Delft (red) and are consistent within the error bars.
The results are in good agreement with experiments showing that these two transport properties that are obtained from the PES are described satisfactorily by the Madrid-Transport force field.

\begin{center}
\begin{figure}[!hbt] \centering
    \centering
    \includegraphics*[clip,scale=0.35,angle=0.0]{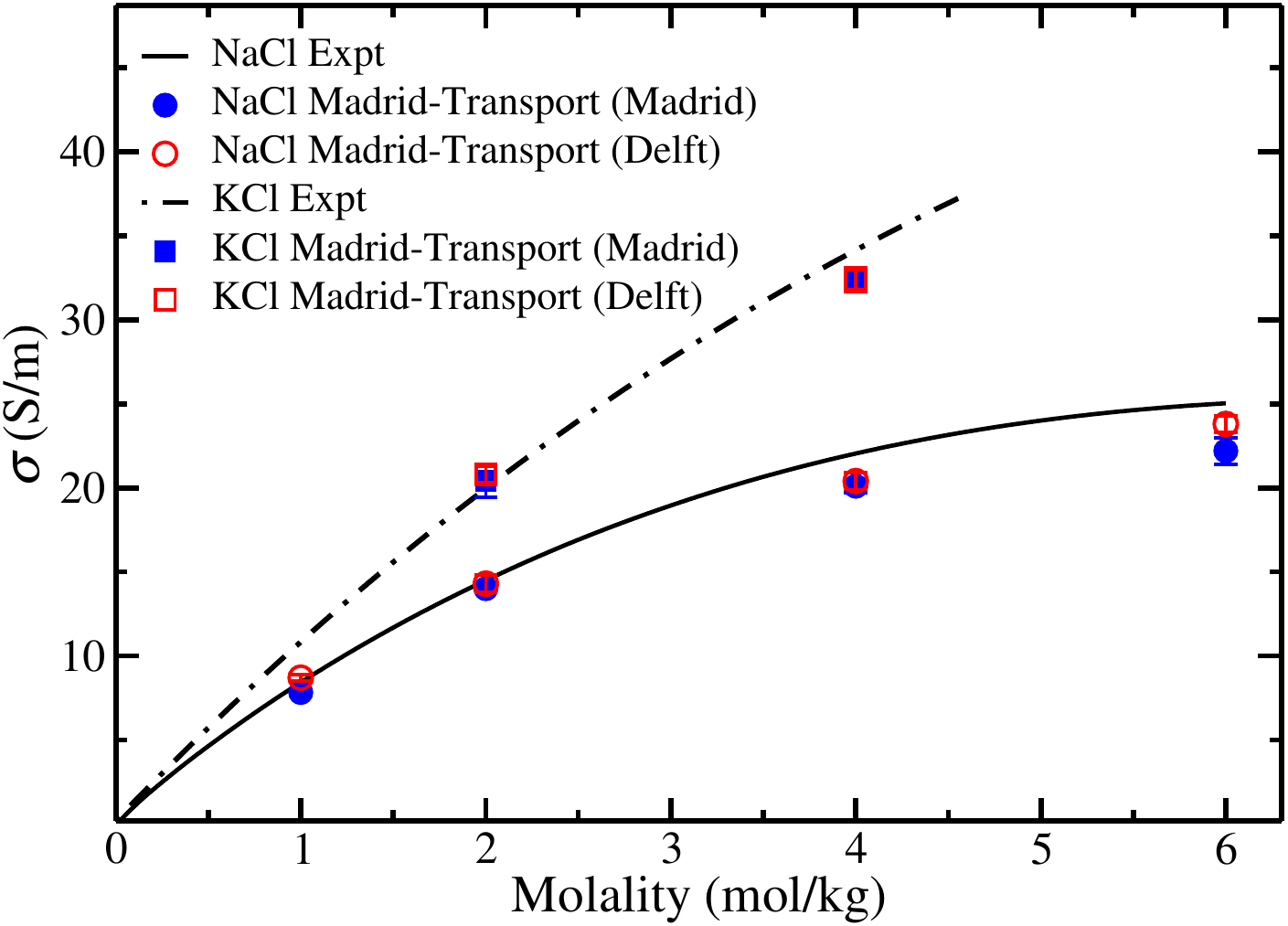}
        \caption{Electrical conductivities as a function of NaCl concentration
	obtained with the Madrid-Transport force field in independent research groups: Madrid (blue) using Eq. \ref{conduc-einstein2} and Delft (red) using Eq. \ref{sigma-delft}, for NaCl (circles) and KCl (squares) aqueous solutions
	at temperature $T$ = 298.15 K, pressure $p$ = 1 bar. Experimental results have been
        taken from Ref.\cite{chambers1956conductances}}
    \label{conductivities-comparado}
\end{figure}
\end{center}

In Fig. \ref{conductivities-comparado} the results for the electrical conductivities of NaCl and KCl solutions using the Madrid-Transport force field are shown. Note
that these results are computed with our novel approach (i.e., using scaled charges ($q$=$\pm0.75$) to describe the trajectories of the system and 
unit charges to compute the electrical conductivities).
We show the results obtained from two different research groups: Madrid (blue) and Delft (red). Both groups have adopted the same approach to calculate conductivities (i.e., employing the EH
equations) with some minor differences. Madrid has evaluated the conductivity
from the mean square dipole displacement, taking into account all the interactions between ions (cation-cation, anion-anion and cation-anion). 
Delft has evaluated 
the Onsager coefficients independently for the cation-cation, anion-anion, and cation-anion interactions, and then summed up all the contributions. 
Both approaches are equivalent and, thus, should yield identical results.
It is important to note that both groups have used different
software, constraint algorithms, system sizes, simulation times, and fitting methods (see Simulation Details). 
Even so, the results for the conductivities obtained by both groups are equal within the error bars. 
Note also that despite the system sizes  used by Delft and Madrid groups are different (between 555 and 4440 water 
molecules), the electrical conductivities are in agreement, showing that no finite size effects in the computation of electrical conductivities are observed.
Nevertheless, since both self and collective (Maxwell Stefan and Fick) diffusivities exhibit significant
finite size effects\cite{Yehhummer,celebi2021finite,jamali2018finite,jamali2020generalized}, a thorough investigation for electric conductivities should be also performed. 
This is particularly important for small concentrations (i.e., below 1 $m$).
This was also previously studied by Shao
et \textit{al.}\cite{shao2020role} for the JC-SPC/E model concluding that there were not finite size effects when using 
the Green-Kubo equation.
In Fig. \ref{conductivities-comparado}, we can  observe that the conductivities obtained by both groups for the Madrid-Transport force field reproduce the experimental conductivities
of both NaCl and KCl aqueous solutions for the whole concentration range. 
Note that the electrical conductivity of KCl is significantly larger than that of NaCl at the same concentration. 
This difference is correctly described by the Madrid-Transport force field. 
Thus, the Madrid-Transport force field and
the use of scaled charges for the trajectories and integer charges for computing conductivities (i.e., describing the PES with scaled charges and the DMS with integer charges)
allows for the first time to correctly reproduce experimental conductivities (and other transport properties as 
viscosities and water diffusion coefficients).
In Tables \ref{tab:NaCl_table} and \ref{tab:KCl_table},  we have collected the computed conductivities for each system (by both groups) along with the conductivities
obtained by using the NE equation (Eq. \ref{nernst-einstein}) with and without  applying the finite-size corrections to the 
self-diffusion coefficients. 

\begin{table*}[hbt!]
	\caption{Computed electrical conductivities ($\sigma$ in units of S$\cdot$m$^{-1}$) of aqueous NaCl solutions from the EH relations (Eq. \ref{conduc-einstein2}) 
	for different molalities ($m$ in units of mol\textsubscript{salt} kg\textsubscript{water}\textsuperscript{-1}) and system sizes. 
	All simulations were performed at 1 bar and 298.15 K, using the Madrid-Transport model. The  number of water molecules ($n_\text{W}$) and NaCl molecules ($n_\text{s}$), 
	the corresponding densities ($\rho$ in units of kg$\,$m$^{-3}$) and viscosities ($\eta$ in units of mPa$\cdot$s) are shown for all molalities. 
	Additional electrical conductivities computed using the Nernst-Einstein with ($\sigma_\text{NE+YH}$ in units of S$\cdot$m$^{-1}$) 
	and without ($\sigma_\text{NE}$ in units of S$\cdot$m$^{-1}$) Yeh-Hummer finite-size corrections\cite{Yehhummer,celebi2021finite,jamali2018shear} are reported as well.
	Numbers in parentheses are the uncertainty in the last digit of the results.}
	\label{tab:NaCl_table}
	\centering
\begin{tabular}{ccccccccc}
	\hline
	\hline
	 & $m$ & $n_\text{w}$ & $n_\text{s}$ & $\rho$ & $\eta$&$\sigma$   &$\sigma_\text{NE}$ & $\sigma_\text{NE+YH}$ \\ 
\hline
Expt. & 0 & - & -  &       997.043                           &  0.89  & 0 & - & - \\
Madrid & 0 & 4440  & 0  &   997.3(3)                           &  0.85(5)  & 0 & 0 & 0\\
Delft & 0 & 1000  & 0  &   997.9(3)                           &  0.82(1)  & 0 & 0 & 0\\
\hline
	Expt & 1 & - & -  &       1036.21                           & 0.97  & 8.48& - & - \\
        Madrid  & 1 & 4440 & 80  &      1035.2(5)                           &  0.97(7)  & 7.8(1) & 9.7(2) & 10.6(2) \\
       Delft &   1  &   1000   &   18    &   1034.8(1) &   0.929(9)     &  8.7(2) &   9.6(1)  &   11.2(1)  \\ 
        Delft &   1  &   555    &   10    &   1034.9(1) &   0.93(1)     &  8.7(2) &   9.6(2)  &   11.5(2)  \\
\hline
	Expt. &   2  & -  & -    &   1072.27           &            1.08         &   14.49      & -                     & -   \\
Madrid  &   2  &   4440   &   160     &   1070.3(5)           &              1.12(7)      &  14.0(1)     &   17.3(1)                    &   18.9(1)    \\
        Delft &   2  &   1000   &   36          &  1069.9(2)  &   1.07(2) &  14.3(5)         &   17.3(2)  &   19.9(2)  \\ 
         Delft &   2  &   555    &   20  &  1070.1(1)  &   1.08(2)  &  13.8(6)        &   16.7(1)  &   19.9(1)  \\
	\hline
	 Expt. &   4  & -  & -  &   1136.91           &      1.35            &   22.04      & -             & -  \\
         Madrid &   4  &   4440   &   320     &   1135.4(5)                 &      1.44(10)                &  20.1(4)     &   27.3(1)            &   29.6(1)    \\
         Delft &   4  &   1000    &   72  &  1134.92(7) &    1.37(3)   &  20.4(8)      &   27.4(2)  &   31.3(2)  \\
         Delft &   4  &   555    &   40  &  1135.31(9) &    1.32(3)   &  21.6(5)      &   26.5(2)  &   31.4(1)  \\
	\hline
	Expt. &   6  & -  & -  &   1192.88           &      1.75            &  25.03       & -             & -  \\
 Madrid &   6  &   4440   &   480   &  1194.5(5)       &             1.79(10)             &  22.2(8)     &   32.6(01) &   35.2(01)    \\
  Delft &   6  &   1000   &   108   &  1194.0(2)   &   1.69(2)  &  23.8(5)    &   32.53(7) &   37.04(8)  \\ 
  Delft &   6  &   555    &   60    &  1194.0(1)   &   1.69(1)  &  23.8(6)    &   32.2(3)  &   37.7(3)  \\
\hline
\hline
\end{tabular}
\end{table*}

\begin{table*}[hbt!]
	\caption{Computed electrical conductivities ($\sigma$ in units of S$\cdot$m$^{-1}$) of aqueous KCl solutions from the EH relations (Eq. \ref{conduc-einstein2})
        for different molalities ($m$ in units of mol\textsubscript{salt} kg\textsubscript{water}\textsuperscript{-1}) and system sizes.
        All simulations were performed at 1 bar and 298.15 K, using the Madrid-Transport model. The  number of water molecules ($n_\text{W}$) and KCl molecules ($n_\text{s}$),
        the corresponding densities ($\rho$ in units of kg$\,$m$^{-3}$) and viscosities ($\eta$ in units of mPa$\cdot$s) are shown for all molalities. 
        Additional electrical conductivities computed using the Nernst-Einstein with ($\sigma_\text{NE+YH}$ in units of S$\cdot$m$^{-1}$)
        and without ($\sigma_\text{NE}$ in units of S$\cdot$m$^{-1}$) Yeh-Hummer finite-size corrections\cite{Yehhummer,celebi2021finite,jamali2018shear} are reported as well.
	        Numbers in parentheses are the uncertainty in the last digit of the results.}
	\label{tab:KCl_table}
	         \centering
\begin{tabular}{ccccccccc}
        \hline
        \hline
         & $m$ & $n_\text{w}$ & $n_\text{s}$ & $\rho$ & $\eta$&$\sigma$   &$\sigma_\text{NE}$ & $\sigma_\text{NE+YH}$ \\
\hline
      Expt. & 0 & - & -  &       997.043                           &  0.89  & 0 & - & - \\
Madrid & 0 & 4440  & 0  &   997.3(3)                           &  0.85(5)  & 0 & 0 & 0\\
Delft & 0 & 1000  & 0  &   997.9(3)                           &  0.82(1)  & 0 & 0 & 0\\
\hline
        Expt. &   2  & - & -  &       1081.5                          &    0.90   & 19.98 & - & - \\
                Madrid     &   2  &   4440        &   160      &  1081.1(5) &    0.95(5)  &  20.4(9)              &   24.0(1)     &   25.8(2)  \\
                Delft     &   2  &   1000     &   36       &  1080.6(1) &   0.91(4)   &  20.8(3)       &   23.7(2)       &   26.7(3)  \\ 
                Delft     &   2  &   555      &   20       &  1080.8(1) &   0.92(2) &  20.8(5)         &   22.8(4)       &   26.5(3)  \\
\hline  
Expt. &   4  & - & -  &       1152.2                          &    0.94   & 34.15 & - & - \\
        Madrid  &   4  &   4440         &    320      &                1152.3(5)          &   1.03(7)  &  32.5(6)        &   40.5(1)    &   43.1(1)    \\
        Delft    &   4  &   1000       &   72       &  1151.60(5) &    1.00(2)   &  32.9(9)        &   39.7(7)   &   44.8(7)  \\
        Delft    &   4  &   555       &   40       &  1151.81(3) &    0.99(2)   &  32.4(7)        &   38.8(2)   &   45.1(3)  \\
\hline
\hline
\end{tabular}
\end{table*}

%We have shown that using scaled charges for trajectories and unit charges for computing conductivities yields to 
%a good description of electrical conductivities. Is this true for all  scaled charges? To answer this question we have evaluated the electrical
%conductivities of the Madrid-2019 force field which assign a charge of q=$\pm=0.85$ for the ions. Applying the same 
%methodology that we have described previously for the Madrid-Transport we proceed now to compute the conductivities 
%of Madrid-2019 force field as we show in Fig. \ref{conductivities-2019}. It is clear that the Madrid-2019 is not able to reproduce the experimental
%conductivities of NaCl solutions. The results demonstrate that not all scaled charges can reproduce electrical conductivities.
%
%
%\begin{center}
%\begin{figure}[!hbt] \centering
%    \centering
%    \includegraphics*[clip,scale=0.35,angle=0.0]{conductividad-2019.eps}
%        \caption{Electrical conductivities as a function of NaCl concentration
%        obtained with the scaled charge models Madrid-2019 and Madrid-Transport force fields using the EH equation at temperature T = 298.15 K, pressure p = 1 bar. 
%Results obtained by Madrid group are showed in blue and those from Delft group in red.
%	Experimental results have been
%        taken from ref.\cite{chambers1956conductances}}
%    \label{conductivities-2019}
%\end{figure}
%\end{center}
%

\subsection{Electrical Conductivities by using the Nernst-Einstein equation}

Although the correct way to calculate conductivities is to use the Green-Kubo or the Einstein-Helfand equations.
However, the NE
equation (Eq. \ref{nernst-einstein}) is widely used due to its simplicity.
In  Tables \ref{tab:NaCl_table} and \ref{tab:KCl_table},  the electrical conductivities
predicted by the approximate NE formula are also presented. 
As can be seen, the 
NE relation overestimates 
the true conductivity of the model as obtained from the EH equations. 
Not surprisingly, the NE equation does not provide the exact value of the electrical conductivity of the model
(see the Supporting Information). 
The reason for that is that it neglects correlations between different ions.
The deviations increase with the concentration but they are clearly visible even at a concentration of 1 $m$.
Our results suggest that correlations between different ions tend to decrease the value of the electrical conductivity.
 The  conclusion from the results of  Tables \ref{tab:NaCl_table} and \ref{tab:KCl_table} is
 that NE should not be used to estimate electrical conductivities as it provides incorrect results. 
If one wants to compute the true conductivity of a force field, either the GK or EH expressions should be used. 

\begin{center}
\begin{figure*}[!hbt] \centering
    \centering
    \includegraphics*[clip,scale=0.31,angle=0.0]{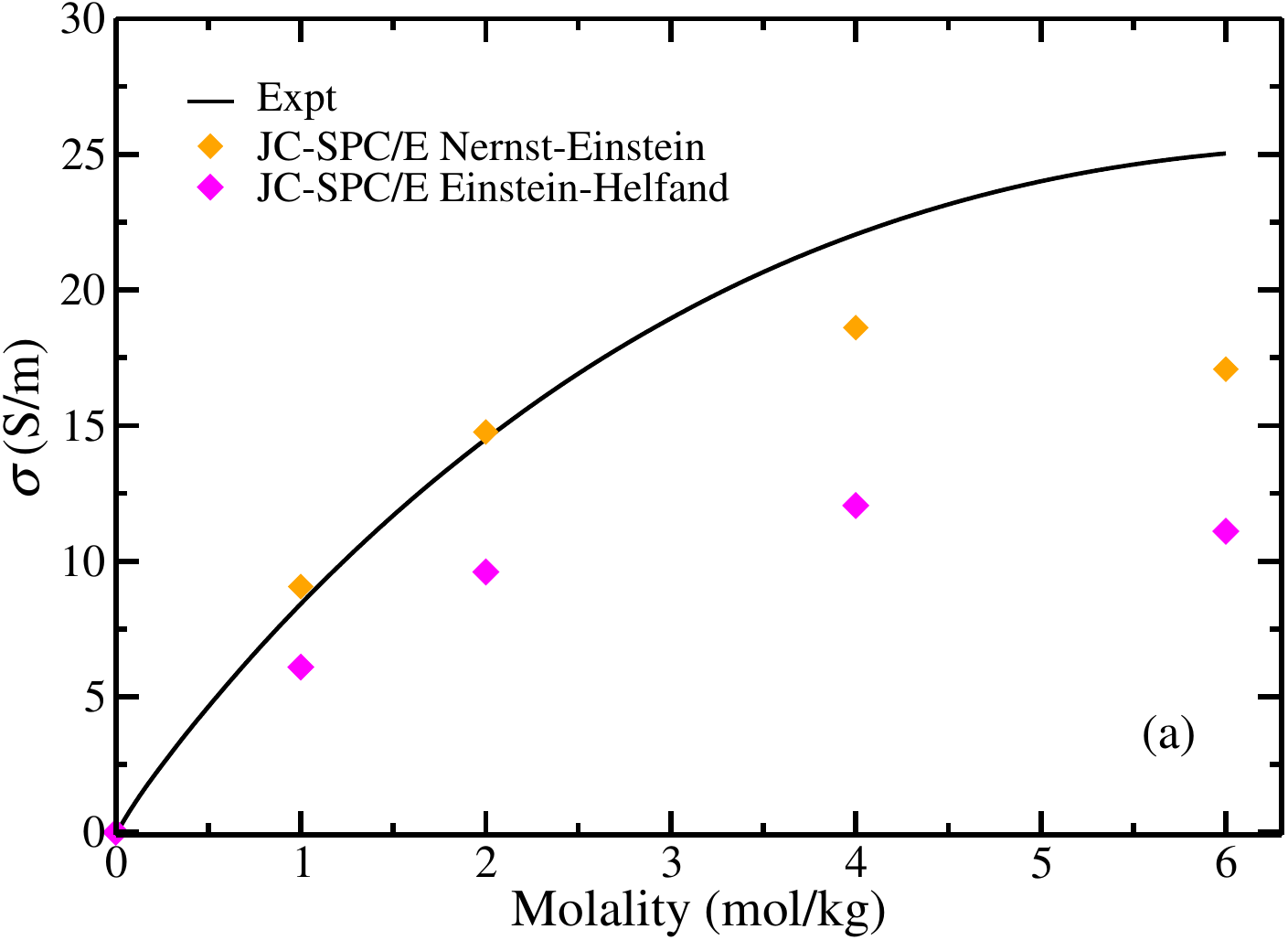}
    \includegraphics*[clip,scale=0.31,angle=0.0]{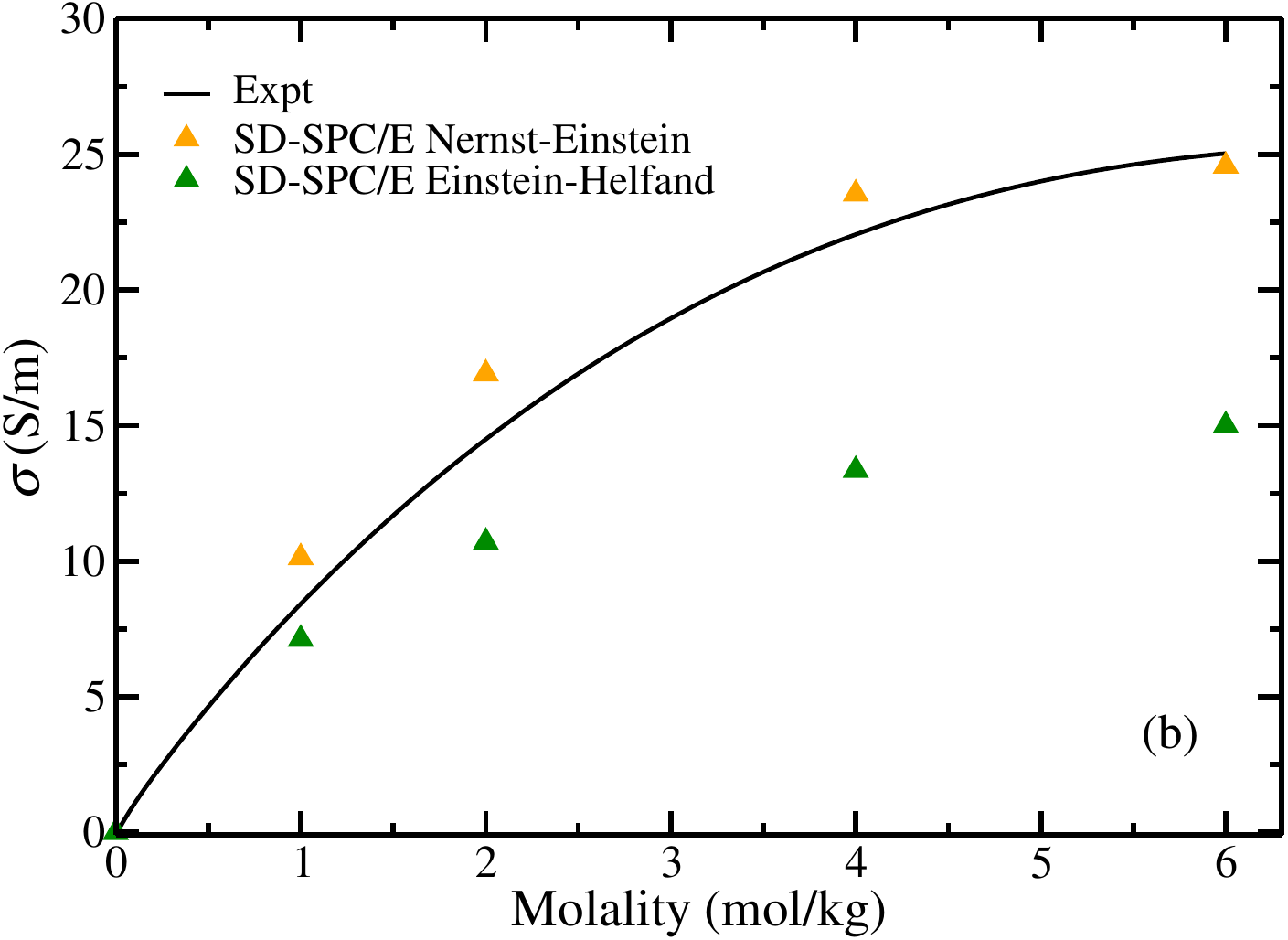}
        \caption{Electrical conductivities as a function of NaCl concentration
       obtained with different models studied in this work by two different methodologies
	(i.e., by using the EH relation and the Nernst-Einstein equation) at $T$ = 298.15 K,  $p$ = 1 bar. (a) JC-SPC/E (b) SD-SPC/E. Experimental results have been
        taken from Ref.\cite{chambers1956conductances}}
    \label{conductivities-NE}
\end{figure*}
\end{center}

Despite being inaccurate, the NE expression is often used in many papers to obtain electrical conductivities.
There are two main reasons for this.
The first one is that the calculation of self-diffusion coefficients is relatively easy and computationally cheap. 
Thus, if the NE formalism is used, the electrical conductivities are obtained with no additional computational cost.
Diffusion coefficients have good statistics as one can accumulate the statistics of each individual ion,
while the EH is expensive as it is a global property and each configuration contributes a single value of the correlation function. 
In short, diffusion coefficients of ions can be obtained with good accuracy in runs of 20 ns, whereas ca. 600 ns 
are needed to have  reasonable statistics of the electrical conductivity.
Additionally, the computation of the diffusion coefficient is implemented in many MD packages, whereas this is not the case 
of the electrical conductivity. 
The second reason why NE is so popular is because most of the force fields tend to underestimate significantly the electrical conductivity
compared to the experiments when computed rigorously from the EH formalism. 
However, since NE electrical conductivities are much larger they tend to be in better agreement
with the experimental results. This
creates the paradox that despite NE tends to overestimate the electric conductivities, in many cases is closer
to experimental data, and thus, there is a
resistance to abandon its use.  Obviously, this apparent agreement arises from a cancellation of errors 
(i.e., a poor force field along with a poor way of computing the actual conductivity of the force field can provide good agreement with experiments). 
This is illustrated in Fig. \ref{conductivities-NE}, where the electrical conductivities of the JC-SPC/E and SD-SPC/E obtained from NE and from EH are compared to experiment. 
As it can be seen, the agreement is better with NE.  As we have shown NE  does not describe correctly the electrical conductivity of the force 
field so that the apparent improvement is obtained from fortuitous cancellation of two errors (the force field and the way to compute the electrical conductivity). 
Another "apparent" advantage of the NE formalism is that since diffusion coefficients are quite sensitive 
to the size of the system, one can often find a system size for which the agreement with experiment is excellent.
This adds another degree of freedom
for "fine-tuning" the final value of electric conductivity.
However, this should not be accepted when striving for accurate computation of properties. 
This becomes even more pronounced by the fact that the system size dependency of the electrical conductivity when computed from the EH formalism is quite small
as it has been shown in this work (i.e., compare the results from Madrid and from Delft) and also shown in other works\cite{shao2020role}. 

\begin{center}
\begin{figure}[!hbt] \centering
    \centering
    \includegraphics*[clip,scale=0.35,angle=0.0]{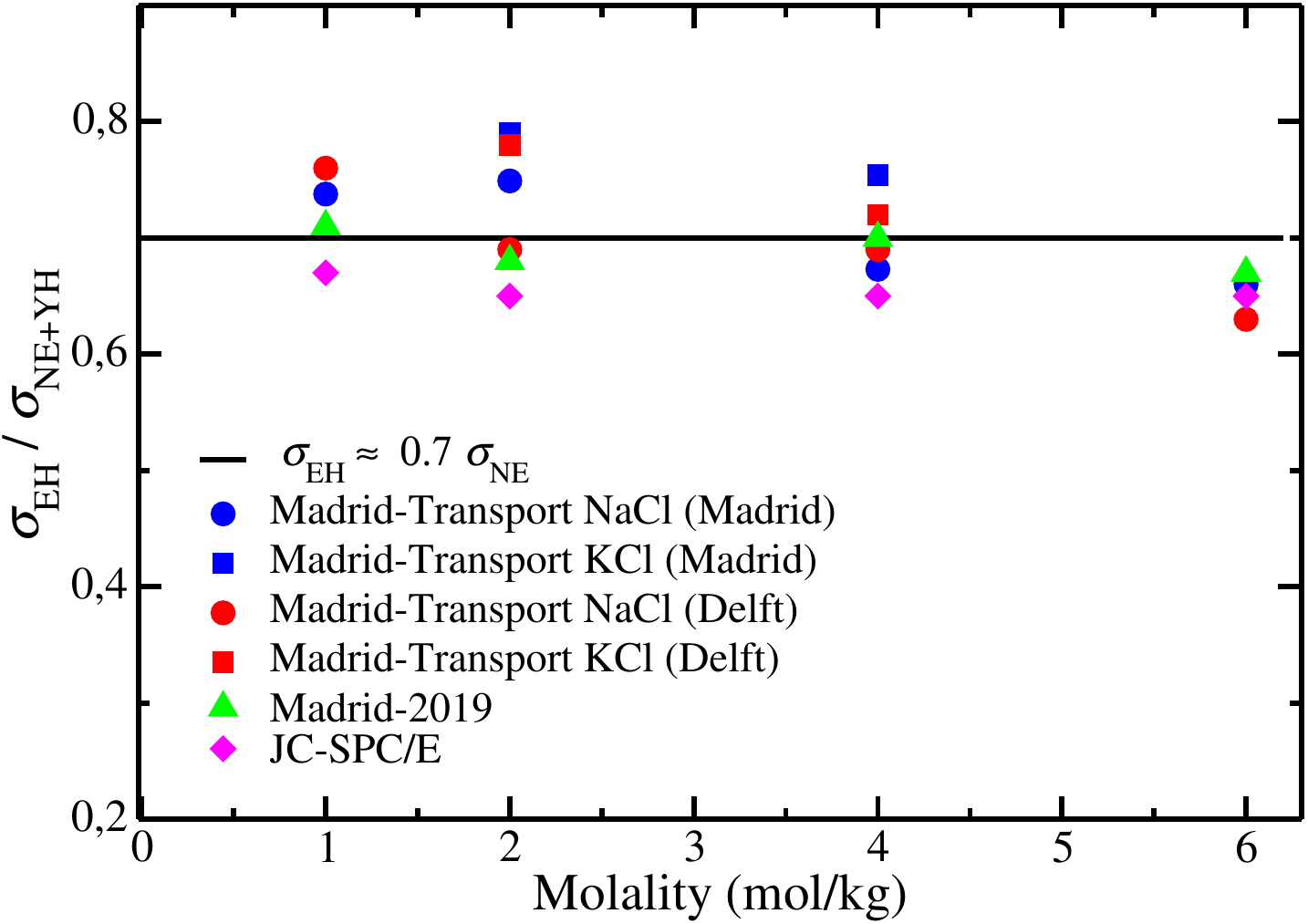}
        \caption{Ratio between the conductivities obtained by using the Einstein-Helfand formula ($\sigma_{EH}$) and the Nernst-Einstein equation ($\sigma_{NE+YH}$).}
    \label{gk-ne-rel}
\end{figure}
\end{center}

After this work,
we strongly advise against using NE when the one can actually compute the conductivity using rigorous ways.
NE can be used when one has no access to such a computation or when a quick estimation of the order of magnitude of the conductivity is needed.
However, even in such cases, the researcher should keep in mind that the NE value is overestimated respect to the real conductivity.
However, we will suggest an approximate 
(and not rigorous) way of at least correcting NE results. 
  The idea is simple. We have analyzed the typical ratio between the electrical conductivities obtained rigorously (i.e., EH
relation) and those obtained from the approximate NE formalism, and have analyzed whether it is approximately constant for different force fields and concentrations.
When computing the electrical conductivity from NE, we have  used the finite size corrected diffusion coefficients.
In this way, the NE conductivities computed here are determined  using one of the most correct approaches to estimate diffusion coefficients in the thermodynamic limit.
In Fig. \ref{gk-ne-rel}, we show the ratio between the conductivities obtained by EH divided by the conductivities calculated with NE as a function
of the concentration. Notice that for models with different charges (JC-SPC/E, Madrid-2019 and Madrid-Transport), for all concentrations and even for both studied salts (NaCl and KCl),
the conductivities using the EH equation are about 30$\%$ lower than when using the NE equation. Thus, a rough approximation of the EH conductivities can be
obtained simply by employing:

\begin{equation}
\label{NE-to-EH}
        \sigma^{EH} \approx 0.7\cdot\sigma^{NE+YH},
\end{equation}
where $\sigma^{EH}$ is the conductivity rigorously computed by using the EH relation, and $\sigma^{NE+YH}$ is the approximate conductivity calculated with the NE equation
(after including Yeh-Hummer corrections to the values of the diffusion coefficients).
If one wants to roughly estimate the correct conductivities of 
a model (i.e., evaluated by the EH relation), one can only employ the NE equation and then apply our rule described in Eq. \ref{NE-to-EH}.
In this way, one can obtain an approximate (with a typical error of about 5-8$\%$)
but still reasonable estimate of the true conductivity of the force field under consideration from the initial guess provided by the NE relation.
In fact, regarding the recent work from Gullbrekken \textit{et al.}\cite{gullbrekken2023charge}, this rule also works properly for their results.
Note also that this scaling factor in Eq. \ref{NE-to-EH} (0.7) is empirical and is not related to the charge of ions in the force field.
In any case, we recommend to evaluate properly the conductivities by using the GK or EH relations without any approximation to obtain rigorously the
correct conductivity of the force field.

%\begin{table*}[hbt!]
%        \caption{ Results for the electrical conductivities 
%        obtained with the different models studied in this work for aqueous solutions of NaCl at temperature T = 298.15 K, pressure p = 1 bar and at different
%concentrations below experimental solubility. Experimental results have been
%        taken from ref.\cite{chambers1956conductances}.
%	\textcolor{red}{Data fit 1 ns}}
%\label{tabla-difusion-na}
%  \begin{center}
%    \begin{tabular}{c c c c c c}
%\hline
%\hline
%Molality             &  \multicolumn{5}{c}{$\sigma$} \\
%	    mol/kg & \multicolumn{5}{c}{mS$\cdot$cm$^{-1}$} \\
%     \cline{2-6}
%               &      & q = $\pm$1    &  q = $\pm$1  & q = $\pm$0.85 & q = $\pm$0.75  \\
%	    m  & Expt & JC-SPC/E      &  SD-SPC/E    & Madrid-2019   &   Madrid-Transport \\
%\hline
%            1  & 84.2 & 60.91 & 71.17  & 65.23 & 71.72 \\
%            2  & 144.9 & 95.99 & 107.23  & 106.66 & 144.10 \\
%            4  & 220.4 & 120.51 & 122.76  & 156.36 & 185.71 \\
%%            5  & 238.6 & - & -  & - & 211.85 \\
%            6  & 250.3 & 110.99 & 148.61  & 163.52 & 243.35 \\
%\hline
%\hline
%    \end{tabular}
%  \end{center}
%\end{table*}

\section{Conclusions}

In this work, we have evaluated the electrical conductivities of NaCl and KCl aqueous solutions up to their solubility limit by using 
the Einstein-Helfand equation.
We have computed the electrical conductivities of four different NaCl force fields (i.e., JC-SPC/E, SD-SPC/E, Madrid-2019 and Madrid-Transport). 
We have shown that force fields with integer charges are not able to reproduce electrical conductivities, 
as they underestimate them considerably.
Note that for these models integer charges are used to describe both the PES and the DMS. 
We tested a new  approach that is based on the more general idea that different charges
should be used to reproduce correctly the PES and the DMS. 
  To implement this idea, we used the Madrid-Transport force field for NaCl and KCl (with scaled charges of $\pm$0.75 for the ions)
  to obtain the trajectories of the system in the absence of an external electric field. 
The EH equation is used with integer charges for the ions. 
The basic idea is that scaled charges describe the PES better, and integer charges describe the DMS more accurately, as was suggested some time ago.\cite{vegamp15,jorge2019dielectric} 
By using this approach, we have shown that the Madrid-Transport force field (a model which provides excellent results for transport
properties such as viscosities and diffusion coefficients) reproduces the electrical conductivities of NaCl and KCl for the whole concentration
range. 
Certainly, one can argue that it is possible to use a model with $q$=$\pm$1 to describe the PES and with $q$
higher than $\pm$1 for describing the DMS. 
By using this idea (developed in this work)  the agreement with experiments for electrical conductivities will also
improve for models that use integer charges for the PES.
However, the cost of that is that the transported charge will not be 1 $e$ , a result that has no physical meaning.
In addition, the force field with that charge would not reproduce the viscosities and self-diffusion coefficients of the water
as in the rest of unit charge models.
 It seems therefore that to reproduce the DMS one should use the integer charge that actually is transported (1 $e$)
 but in contrast, for describing the PES, one can use scaled charges to better describe the relative weight of some configurations with respect to others.
What quantum chemistry says about the value of the charge that is indeed transported? Is it correct to assume  formal integer 
charges when computing electrical conductivities? This issue has been discussed in two important papers\cite{grasselli2019topological,french2011dynamical}.
For simplicity we shall discuss this issue using the Green-Kubo formalism\cite{hansen2013theory,bacstuug2005temperature} but one could also use the totally equivalent Helfand-Einstein relation. 
The electrictical conductivity can be rigorously computed from ab-initio calculations using the expression:

\begin{equation}
\label{conduc-GK}
        \sigma = \frac{1}{3Vk_{B}T}  \int_{0}^{\infty} \Big \langle \Vec{j}_{i}(t)\cdot\Vec{j}_{i}(0) \Big \rangle dt ,
\end{equation}

where $\Vec{j}_{i}$ is the current density vector which is defined as:

\begin{equation}
\label{current}
	\Vec{j}_{i}(t)=\sum_{i}^{N} {\bf Q}^{*}_{i}(t) \cdot \Vec{v}_{i}(t),
\end{equation}

Apparently everything seems normal, but now comes the first surprise. The charge {\bf Q}$^{*}_{i}$(t) is not a scalar but a time dependent tensor with components
given by:  
\begin{equation}
	Q^{*}_{i;\alpha,\beta} =\frac{\partial M_{\alpha}}{\partial r_{i,\beta}}
\end{equation}
where $\alpha,\beta=x,y,z$ and 
r$_{i,\beta}$ refers to the component $\beta$ of the vector defining the position of ion $i$ and $M_{\alpha}$ is the $\alpha$ component of 
the dipole moment of the system. 
It should not come as a surprise that the value of the charge of the ion for the calculation of the conductivity 
comes from derivatives of the dipole moment of the system which 
is well defined and not from arbitrary schemes partitioning the electronic density among the different ions of the system (as for 
instance the Bader method\cite{bader1990quantum,sanville2007improved}). 
   However, it has been shown\cite{grasselli2019topological,french2011dynamical} that one can obtain the rigorous value of the conductivities 
using a non-time dependent scalar for the charge of ion $i$ (usually denoted as the topological charge $q_{i,top}$).    
This is summarized in the following equations:
\begin{equation}
\label{current-topological}
       \Vec{j'}_{i}(t)=\sum_{i}^{N} q_{i,top} \cdot \Vec{v}_{i}(t),
\end{equation}
\begin{equation}
\label{conduc-GK-equals}
        \int_{0}^{\infty} \Big \langle \Vec{j}_{i}(t)\cdot\Vec{j}_{i}(0) \Big \rangle dt = \int_{0}^{\infty} \Big \langle \Vec{j'}_{i}(t)\cdot\Vec{j'}_{i}(0) \Big \rangle dt
\end{equation}

The conclusion is that although the charge of an ion is a time dependent tensor, there is a non-time dependent scalar value of the charge (denoted as the topological charge) 
  that leads to the correct value of the electrical conductivity. Finally, Grasselli and Baroni\cite{grasselli2019topological}, and French et \textit{al.}\cite{french2011dynamical} 
  have shown that the value of the topological charge is just the value of the formal charge. Quantum chemistry 
  suports the use of the formal integer charge (i.e., 1 $e$ for monovalent ions) in the calculation of the 
  electrical conductivity. One should not use different values of the formal charge when computing electrical conductivities. That is what we have done in this work. 
  However, quantum chemistry does not say anything about the charge that better fits the PES as it should be regarded as a fitting parameter of the force field, and we have shown 
  that for  aqueous electrolyte solutions the scaled charge with value 0.75 $e$ provides an excellent description of transport properties.

Finally, we have shown that when using the Nernst-Einstein equation the electrical conductivities are incorrectly calculated due to the neglect of
the correlation between different ions,
showing discrepancies even at low concentrations. We propose a rule of thumb to obtain a rough estimate of the EH conductivities.
The recipe is simple. It consists in  multiplying by  0.7 the conductivity obtained from the NE equation 
(after applying YH corrections to the values of the diffusion coefficients of the individual ions). 
Nevertheless, this is an approximate correction, and our advice is to use the correct expression
(i.e., EH or GK) to obtain rigorously the electrical conductivity of a certain force field. 

The results presented in this work independently obtained by two research groups could be useful in the future as benchmark
results to be reproduced by groups interested in computing electrical conductivities of electrolyte solutions.
    The success of the Madrid-Transport force field in reproducing the transport properties of NaCl and KCl solutions
    can be regarded as workcase example showing how fruitful the idea of using different charges to describe the PES and DMS can be in the future.
The community performing classical simulations should benefit from this "mental" flexibility.
In fact, the community performing ab-initio calculations is already using in a effective way as they are using different 
fitting parameters when developing neural networks for the PES and the DMS\cite{krishnamoorthy2021dielectric,piaggi2022homogeneous}. 
Why should we not realize that we can do the same with our force fields?

\section{Conflict of Interest}
The authors have no conflicts to disclose.

\section{Data Availability}
The data that support the findings of this study are available
within the article and in the SI.

 \section{Acknowledgments}
This project has been funded by grant PID2019-105898GB-C21 of the MICINN.
This work was sponsored by NWO Domain Science for the use of super-computer facilities.
O. A. Moultos gratefully acknowledges the support of NVIDIA Corporation with the donation of the Titan V GPU used for this research.

\providecommand{\latin}[1]{#1}
\makeatletter
\providecommand{\doi}
  {\begingroup\let\do\@makeother\dospecials
  \catcode`\{=1 \catcode`\}=2 \doi@aux}
\providecommand{\doi@aux}[1]{\endgroup\texttt{#1}}
\makeatother
\providecommand*\mcitethebibliography{\thebibliography}
\csname @ifundefined\endcsname{endmcitethebibliography}
  {\let\endmcitethebibliography\endthebibliography}{}

\end{document}